\documentclass[aps,prl,floatfix,twocolumn,reprint,superscriptaddress]{revtex4-1}
\usepackage{amsfonts}
\usepackage{mathrsfs}
\usepackage{amsmath}
\usepackage{color}
\usepackage{graphicx}
\usepackage{bm}
\usepackage{amssymb}
\usepackage{xspace}
\usepackage{epstopdf}
\usepackage{dcolumn}
\usepackage{longtable}
\usepackage{multirow}
\usepackage[colorlinks=true, letterpaper=true, pdfstartview=FitV, linkcolor=blue, citecolor=blue, urlcolor=blue]{hyperref}

\newcommand{\sgn}{\text{sgn}}

\makeatletter

\newcommand{\Rmnum}[1]{\expandafter\@slowromancap\romannumeral #1@}
\makeatother

\begin{document}
\title{Chirality Hall Effect in Weyl Semimetals}

\author{Shengyuan A. Yang}
\affiliation{Research Laboratory for Quantum Materials and EPD Pillar, Singapore University of Technology and Design, Singapore 487372, Singapore}

\author{Hui Pan}
\affiliation{Department of Physics, Beihang University, Beijing 100191, China}

\author{Fan Zhang}\email{zhang@utdallas.edu}
\affiliation{Department of Physics, University of Texas at Dallas, Richardson, Texas 75080, USA}

\begin{abstract}
We generalize a semiclassical theory and use the argument of angular momentum conservation
to examine the ballistic transport in lightly-doped Weyl semimetals,
taking into account various phase-space Berry curvatures.
We predict universal transverse shifts of the wave-packet center in transmission and reflection,
perpendicular to the direction in which the Fermi energy or velocities change adiabatically.
The anomalous shifts are opposite for electrons with different chirality,
and can be made imbalanced by breaking inversion symmetry.
We discuss how to utilize local gates, strain effects, and circularly polarized lights
to generate and probe such a chirality Hall effect.
\end{abstract}
\maketitle

When a strong topological insulator~\cite{Kane,Qi} undergoes a phase transition to a trivial band insulator in three dimensions,
a gapless Dirac point~\cite{Murakami,Young,Fang,Na} emerges at the critical point,
if both time-reversal ($\mathcal{T}$) and inversion ($\mathcal{P}$) symmetries are present.
Remarkably, when one of the two symmetries is broken, the critical point expands in the phase diagram
and the Dirac point splits into pairs of Weyl points related by the unbroken symmetry.
This emergent phase~\cite{Nutrino-Lattice,Volovik,Wan,Balents,Ran,Xu,Bernevig,Lu,Gong,Sau,Das,Yong1,Yang,Liu,Yong2},
dubbed as Weyl semimetal (WSM), is an appealing topological state of matter,
with the Fermi surface being those Weyl points.

In the simplest case when $\mathcal{T}$ symmetry is broken, a WSM at long wavelength only has one pair of Weyl points,
which may be described by the Hamiltonian
\begin{eqnarray}
\mathcal{H}=\tau[v_x k_x \sigma_x + v_y k_y \sigma_y + v_z (k_z-\tau b) \sigma_z + b_0]\;.
\label{Heff}
\end{eqnarray}
Here $v$'s are the Fermi velocities, $\sigma$'s are Pauli matrices,
and $\tau=\pm$ denote the left- and right-handed Weyl fermions,
which are required to come in pairs by the Nielsen-Ninomiya theorem~\cite{no-go}.
$\tau b\hat{z}$ are the positions of the pair of Weyl points in the Brillouin zone (BZ),
and $2b_0$ is their energy splitting, which vanishes when $\mathcal{P}$ symmetry is not broken.
Since all three $\sigma$'s are used up locally at each Weyl point,
small perturbations may renormalize the parameters but cannot open a gap.
Indeed, each Weyl point is protected by the Chern number ($\pm 1$) of the valence band on a constant-energy surface enclosing it.
Weyl points can only be annihilated in pairs of opposite chirality, when they are brought together ($b,b_0\!=\!0$)
or when the translational symmetry is broken by strong interactions or by short-range scatterers.

The topological properties of WSM can be best seen by considering a slice of the BZ normal to $\hat{z}$.
The Chern number of the slice changes from $0$ to $1$ and back to $0$ as it crosses the two Weyl points successively.
As a consequence, each nontrivial slice contributes an $e^2/h$ to the Hall conductivity~\cite{Wan,Balents,Ran}
producing $\sigma_{xy}=b e^2/\pi h$ in the bulk,
and also contributes one edge state to the surface
forming a surface Fermi arc connecting the two projected Weyl points.
Recently, the Weyl points and surface arcs appear to be observed in optical experiments~\cite{ex1,ex2,ex3}.
This progress may herald a flurry of exciting experiments on
the appealing transport effects~\cite{AV,Fiete,Pesin} predicted in WSM,
e.g., the chiral magnetic effect~\cite{Aji,Son1,Zyu,Gru,Gos,Hos,Franz,Burkov-14}
when $b_0$ becomes nontrivial in the absence of $\mathcal{P}$ symmetry,
and the axial magnetic effect~\cite{CX,Voz} when $\tau b$,
viewed as a gauge field coupling oppositely to the left- and right-handed Weyl fermions, varies spatially.

Here we discover a universal effect in lightly-doped WSMs,
by examining the semiclassical dynamics of Weyl quasiparticles in the ballistic regime.
The presence of Weyl points leads to substantial Berry curvatures.
Under a longitudinal force field,
the vector product of the force and the curvature results in a transverse velocity of the Weyl wave-packet,
i.e., a velocity counterpart of the Lorentz force in the semiclassical description.
Therefore, as the Weyl wave-packet propagates in the WSM, its center acquires a transverse shift upon transmission and reflection,
perpendicular to the direction in which the Fermi energy or velocities change.
The anomalous shifts are opposite for Weyl fermions with different chirality,
and can be made imbalanced by breaking $\mathcal{P}$ symmetry.
This effect has no counterpart in 2D and is analogous to the Imbert-Fedorov
shift~\cite{Fedorov,Imbert,Onoda,Bliokh,Hosten,Yin} observed for a circularly polarized light beam,
a bosonic cousin of the Weyl wave-packet, due to variation of dielectric constant.

We focus on the simple case of model~(\ref{Heff}) with $b_0\!=\!0$ unless $\mathcal{P}$ symmetry breaking is noticed.
Before the more general discussion based on the semiclassical equations of motion (EOM),
we first derive our main results using the conservation of angular momentum $J_z$,
in the presence of rotational symmetry along $\hat z$, which requires $v_x=v_y$ and isotropic transverse perturbations.
The total angular momentum of a Weyl wave-packet is given by
\begin{eqnarray}
{\bm J}={\bm r}_c\times{\bm k}_c+\frac{\tau}{2}\,{\bm n}\,,
\end{eqnarray}
where $(\bm r_c,\bm k_c)$ is the phase-space center of the wave-packet,
$\bm n$ is the unit vector $(v_xk_x,v_yk_y,v_zk_z)/\mathcal{E}_{\bm k}$,
and $\mathcal{E}_{\bm k}\!=\!\eta\sqrt{v_x^2k_x^2\!+\!v_y^2k_y^2\!+\!v_z^2k_z^2}$
are the energy dispersions of the electron and hole ($\eta=\pm$) bands.
For simplicity, the subscript $c$ will be dropped wherever appropriate.
The first term in $\bm J$ is the orbital angular momentum of the wave-packet,
whereas the second term is the intrinsic $1/2$-pseudospin angular momentum of a Weyl fermion.

\begin{figure}[t!]
\includegraphics[width=8.5cm]{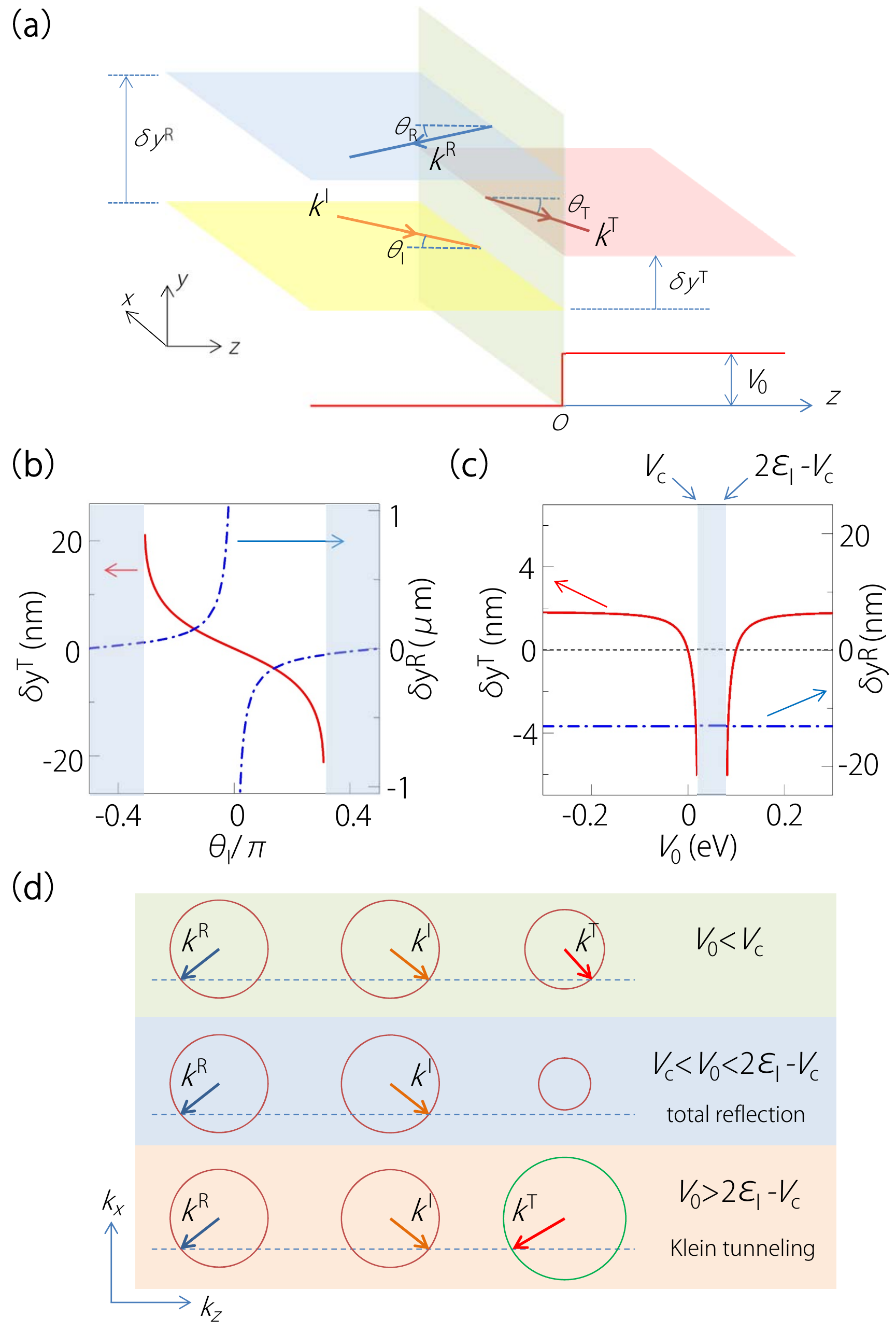}
\caption{(a) Schematic figure showing the transverse shifts $\delta y^\text{T,R}$ in transmission and refection for an incident Weyl electron wave-packet in the $x$-$z$ plane scattered by a potential step.
(b) and (c) $\delta y^\text{T,R}$ plotted as functions of the incident angle $\theta_\text{I}$ and the potential step height $V_0$, respectively.
The shaded regions mark the range of total reflection.
(d) Schematic figure showing the Fermi surfaces and momenta ($k_y=0$) of the reflected and/or transmitted wave-packets in normal, total reflection, and Klein tunneling regimes.
The behaviors in (a)-(c) are for left-handed Weyl electrons whereas the behaviors for right-handed ones (not shown) are opposite; the features in (d) do not distinguish chirality.
We have chosen the parameter values as $\mathcal{E}_\text{I}=0.01$~eV, $V_0=0.014$~eV, $v_z=10^6$~m/s, and $v_{x,y}=3\times 10^5$~m/s in (b),
and $\theta_\text{I}=\pi/4$, $\mathcal{E}_\text{I}=0.05$~eV, $v_z=10^6$~m/s, and $v_{x,y}=8\times 10^5$~m/s in (c).}
\label{fig1}
\end{figure}

We now consider the situation of a Weyl electron wave-packet impinging upon a potential step. As sketched in Fig.~\ref{fig1}(a),
consider an incident electron that has an energy $\mathcal{E}_\text{I}>0$
and an incident angle $\theta_\text{I}=\arctan(k_x^\text{I}/k_z^\text{I})$ in the $x$-$z$ plane.
The potential step satisfies $V=0$ for $z<0$ and $V=V_0$ for $z>0$ in our long-wavelength model,
whereas it is smooth enough at the lattice scale.
All energy scales are assumed to be small compared to the bandwidth
such that the velocities change little across the step.

For $V_0<\mathcal{E}_\text{I}$, the transmitted particle ($\text{T}$) is also a Weyl electron.
The energy and momentum conservations require that $\mathcal{E}_\text{I}=V_0+\mathcal{E}_\text{T}$ and $k_x^\text{I}=k_x^\text{T}$.
The conservation of $J_z$ further determines the trajectory of the transmitted electron,
regardless of the potential and scattering characteristics around $z=0$.
Remarkably, we find that in transmission the wave-packet center acquires a translation perpendicular to the incident plane, given by
\begin{equation}\label{T-symm}
\delta y^\text{T}=\frac{\tau(n_z^\text{T}-n_z^\text{I})}{2k_x^\text{T}}=\frac{\tau v_z}{2}
\left(\frac{\cot\theta_\text{T}}{\mathcal{E}_\text{I}-V_0}-\frac{\cot\theta_\text{I}}{\mathcal{E}_\text{I}}\right),
\end{equation}
where $\theta_\text{T}\!=\!\arctan(k_x^\text{T}/k_z^\text{T})$ is the refraction angle.
Since $v_zk_z^\text{T}\!=\!\sqrt{(\mathcal{E}_\text{I}-V_0)^2-v_x^2{k_x^\text{I}}^2}$,
when $V_0>V_c\equiv\mathcal{E}_\text{I}- v_x |k_x^\text{I}|$,
$k_z^\text{T}$ becomes imaginary and a total reflection occurs.

For $V_0\!>\!\mathcal{E}_\text{I}$, the potential step forms a $p$-$n$ junction and Klein tunneling of the Weyl electron may occur.
The energy and momentum conservations require that $k_x^\text{T}\!=\!k_x^\text{I}$
and $v_zk_z^\text{T}\!=\!-\sqrt{(\mathcal{E}_\text{I}-V_0)^2-v_x^2{k_x^\text{I}}^2}$.
Evidently, when $\mathcal{E}_\text{I}<V_0<2\mathcal{E}_\text{I}-V_c$,
$k_z^\text{T}$ is imaginary and the total reflection occurs.
Yet, when $V_0>2\mathcal{E}_\text{I}-V_c$, Klein tunneling is possible and
the minus sign in $k_z^\text{T}$ means a positive group velocity of the outgoing hole.
Remarkably, the refraction index is negative, i.e, $\theta^\text{I}\theta^\text{T}<0$.
The transverse shift in this case takes the same form of Eq.~(\ref{T-symm}).

An anomalous transverse shift exists in reflection ($\text{R}$) as well.
The conservation of $J_z$ directly leads to the transverse shift of the reflection plane by
\begin{equation}\label{R-symm}
\delta y^\text{R}=\frac{\tau(n_z^\text{R}-n_z^\text{I})}{2k_x^\text{R}}=-\frac{\tau v_z\cot\theta_\text{I}}{\mathcal{E}_\text{I}}\;,
\end{equation}
where $k_x^\text{R}=k_x^\text{I}$ and $k_z^\text{R}=-k_z^\text{I}$.

One observes that for an incident Weyl electron with $\tau=1$,
as shown in Fig.~\ref{fig1}(b), $\delta y^\text{R}$ and $\delta y^\text{T}$ have the same sign that switches at $\theta_\text{I}=0$.
$\delta y^\text{T}$ becomes peaked near total reflection.
Approaching normal incidence, $\delta y^\text{T}$ vanishes whereas $\delta y^\text{R}$ becomes divergent.
The latter anomaly is resolved by the fact that the reflection probability is exponentially small because of the pseudospin conservation.
(Both reflection and transmission probabilities, determined by quantum mechanics, depend on $V_0$ crucially.)
As shown in Fig.~\ref{fig1}(c), $\delta y^\text{R}$ is independent of $V_0$ whereas $\delta y^\text{T}$ strongly depends on $V_0$.
$\delta y^\text{T}$ is positive for $V_0<0$ and negative for $0<V_0<V_c$.
As $V_0$ is further increased, the total reflection occurs until $V_0=2\mathcal{E}_\text{I}-V_c$,
when the Klein tunneling begins to take place and then $\delta y^\text{T}$ has the same sign as $(V_0-2\mathcal{E}_\text{I})$.

Physically, the shape of the Fermi surface changes across the potential step, as sketched in Fig.~\ref{fig1}(d).
Although $k_x$ is conserved, $k_z$ and hence the pseudospin angular momentum $\tau{ n}_z/2$ changes during reflection or transmission.
Therefore, the orbital motion must have a transverse shift to compensate this change, in order to guarantee the conservation of $J_z$.
As suggested by Eqs.~(\ref{T-symm}) and~(\ref{R-symm}),
the transverse shifts are opposite for Weyl electrons with different chirality hence may be dubbed as a chirality Hall effect (CHE).

While the symmetry argument based on $J_z$ conservation can be applied to an interface that is sharp compared to the Fermi wavelength,
the semiclassical description of EOM is more general, regardless of any symmetry.
Generalizing this formalism~\cite{Niu,XCN,Son2,Zhou} to the lightly-doped WSMs,
we find the following velocity and force EOM for the Weyl wave-packet center $(\bm r,\bm k)$:
\begin{eqnarray}
\dot{\bm r}&=&\partial_{\bm k}\mathcal{E}-{\mathit\Omega}_{\bm k\bm r}\cdot \dot{\bm r}-\dot{\bm k}\times\bm{\mathit \Omega}\;,
\label{rdot}\\
\dot{\bm k}&=&-\partial_{\bm r}V-\partial_{\bm r}\mathcal{E}+\mathit{\Omega}_{\bm r\bm k}\cdot\dot{\bm k}\;.
\label{kdot}
\end{eqnarray}
We have included two possible spatially varying ingredients: the electrostatic potential $V$ and the Fermi velocities $v_i$.
The latter not only gives rise to the spatial dependence of the wave-packet energy $\mathcal{E}$,
but also leads to the phase-space Berry curvatures $\mathit\Omega_{\bm k\bm r}$ for the local eigenstates,
where $\mathit\Omega_{k_i r_j}\!=\!\partial_{k_i}\mathcal{A}_{r_j}\!-\!\partial_{r_j}\mathcal{A}_{k_i}$
and $\mathcal{A}_{q_i}\!=\!i\langle u|\partial_{q_i}u\rangle$ (${\bm q}\!=\!{\bm r},{\bm k}$).
In Eq.~(\ref{rdot}) $\bm{\mathit \Omega}$ is the more familiar momentum-space Berry curvature with $\mathit\Omega_\ell\!=\!\epsilon^{ij\ell}\mathit\Omega_{k_i k_j}/2$,
and for the WSM described by model~(\ref{Heff}) $\bm{\mathit\Omega}\!=\!-\tau v_xv_yv_z\bm k/(2\mathcal{E}_{\bm k}^3)$.
The validity of the semiclassical description requires $V$ and $v$'s to be slowly varying spatially compared to the Fermi wavelength.
In this EOM we have considered corrections up to first-order in spatial gradients.

In the presence of rotational symmetry along $\hat z$, the EOM~(\ref{rdot}) and~(\ref{kdot}) are simplified to
$\dot{\bm k}\propto\hat z$ and
$\dot{\bm r}\!=\!\partial_{\bm k}\mathcal{E}-{\mathit\Omega}_{\bm k z}\dot{z}-\dot{\bm k}\times\bm{\mathit\Omega}$,
where ${\mathit\Omega}_{\bm k z}\!=\!-\eta(\partial_{\bm k}\phi)(\partial_z\cos\theta)/2$
and $\theta$ ($\phi$) is the polar (azimuth) angle of $\bm n$.
A straightforward calculation confirms that $\dot{J}_z=0$.
This demonstration shows that the various Berry curvature terms in the EOM are crucial for ensuring the conservation of $J_z$,
and thus for understanding the origin of CHE.

When the rotational symmetry is broken, the argument of $J_z$ conservation is not applicable.
However, one can still use the EOM to obtain the transverse shifts for the same potential step.
The potential step produces a force $\dot{\bm k}\!=\!-\partial_{z}V\hat{z}$ and the Weyl point leads to a substantial curvature $\bm{\mathit\Omega}$.
As seen in Eq.~(\ref{rdot}), these two effects result in an anomalous velocity $-\dot{\bm k}\times\bm{\mathit\Omega}$ of the wave-packet center perpendicular to $\hat z$.
This anomalous velocity is analogous to the Lorentz force in the semiclassical picture.
By integrating Eq.~(\ref{rdot}), we find an anomalous shift
\begin{equation}\label{RT-semi}
\!\!\delta{\bm r}^{\alpha}=-\int_\text{I}^{\alpha}\dot{\bm k}\times{\bm{\mathit\Omega}}\,dt
=\frac{\tau v_x v_y(n_z^\alpha-n_z^\text{I})}{2(v_x^2 k_x^2+v_y^2 k_y^2)}(-k_y, k_x, 0),
\end{equation}
where $\alpha\!=\!\text{T}$ for $V_0<V_c\equiv\mathcal{E}_\text{I}-\sqrt{v_x^2k_x^2+v_y^2 k_y^2}$ and $\alpha\!=\!\text{R}$ for $V_0>V_c$.
Apparently, Eq.~(\ref{RT-semi}) produces exactly the same results as Eqs.~(\ref{T-symm}) and~(\ref{R-symm})
in the case when $v_x=v_y$ and $V_0<\mathcal{E}_\text{I}$.
We note, however, two important issues for this semiclassical result of CHE.
First, the semiclassical trajectory is unique, i.e., transmission if $V_0<V_c$ and reflection if $V_0>V_c$.
Second, Eq.~(\ref{RT-semi}) cannot be applied to the Klein tunneling case,
because the place where $\mathcal{E}_{\bm k}=0$ requires a non-Abelian treatment~\cite{XCN},
which is beyond the scope here.

When $\mathcal{P}$ symmetry is broken, the two Weyl points with opposite chirality would split in energy, as denoted by $\tau b_0$ in model~(\ref{Heff}).
In this case, all above results can still be applied
by replacing $\mathcal{E}_\alpha$ ($\alpha=\text{I},\text{R},\text{T}$) by $\mathcal{E}_\alpha-\tau b_0$.
The imbalanced populations for the left and right chirality lead to a charge Hall effect.
It is even possible to filter one chirality in transmission using the total reflection.

A spatial variation of the Fermi velocities $v$'s can also lead to a CHE for Weyl wave-packets.
Consider an interface around which $v$'s vary along $\hat z$.
In the presence of rotational symmetry, the conservation of $J_z$ dictates that
\begin{equation}\label{v-symm}
\delta y^{\alpha}=\frac{\tau(n_z^{\alpha}-n_z^\text{I})}{2k_x^\text{I}}=\frac{\tau}{2\mathcal{E}_\text{I}}
\left(v_z^\alpha\cot\theta_{\alpha}-v_z^\text{I}\cot\theta_\text{I}\right)\;.
\end{equation}
Interestingly, the total reflection would occur if $({v_x^\text{T}})^2>({v_x^\text{I}})^2+(v_z^\text{I}\cot\theta_\text{I})^2$.
In the absence of rotational symmetry, one can still apply the EOM, assuming $v$'s to be slowly varying.
For the outgoing electron, the anomalous shift in the $x$-$y$ plane can be expressed as
\begin{equation}\label{v-semi}
\delta{\bm r}^\alpha=\int_\text{I}^{\alpha}\frac{-1}{1+\mathit\Omega_{k_z z}}\left[\mathit\Omega_{{\bm k} z}\partial_{k_z}\mathcal{E}
+(\bm{\mathit\Omega}\times\hat{z})\partial_z\mathcal{E}\right]dt\;,
\end{equation}
where whether $\alpha=\text{R}$ or $\text{T}$ is definite and determined by the EOM, as in the case of Eq.~(\ref{RT-semi}).
One observes that there is an additional transverse shift entirely due to $\mathit\Omega_{\bm{k r}}$,
apart from the shift due to the $-\dot{\bm k}\times\bm{\mathit\Omega}$ contribution that has been discussed above.
Compared to momentum-space curvatures $\bm{\mathit\Omega}$,
the phase-space curvatures $\mathit\Omega_{\bm{r k}}$ are less well-known.
Our result in Eq.~(\ref{v-semi}) identifies a physical effect induced by $\mathit\Omega_{\bm{k r}}$.

The velocity-variation-induced CHE is similar to the Hall effect observed for circularly polarized light beams,
i.e., Imbert-Fedorov shifts~\cite{Fedorov,Imbert,Onoda,Bliokh,Hosten,Yin} due to the gradient of dielectric constant in isotropic media~\cite{Onoda} in which $\mathit\Omega_{\bm k\bm r}=0$.
Our result implies that $\mathit\Omega_{\bm k\bm r}$ can also be responsible for a Hall effect of light beams in anisotropic media.
Like photons, the Weyl fermions in WSM also have gapless and linear low-energy dispersions.
However, their pseudospin is $1/2$, the Lorentz invariance is broken at the lattice scale,
their energies can be negative allowing Klein tunneling,
and the separation of $\tau=\pm$ Weyl points in momentum is required to avoid mutual annihilation.
In the reflection of close-to-normal incidence, the Imbert-Fedorov shift vanishes since the spin of light $\tau{\bm n}$ remains the same,
whereas the CHE becomes anomalously large, as we have explained, since the pseudospin of Weyl electron $\tau{\bm n}/2$ becomes opposite.
As great advantages due to their fermionic nature,
gate voltages can easily modify the energetic landscape that Weyl fermions travel through;
breaking $\mathcal{P}$ symmetry can easily create imbalanced populations of the two Weyl flavors.
Interestingly, for photons vacuum is a medium with the largest velocity,
whereas for electrons vacuum can be viewed as a WSM with
the strongest ``intervalley scattering'' leading to an infinite energy gap.
However, as long as the surface Fermi arc appears,
the CHE is anticipated to survive at a WSM-vacuum interface, with a total reflection.

\begin{figure}[t!]
\includegraphics[width=9.0cm]{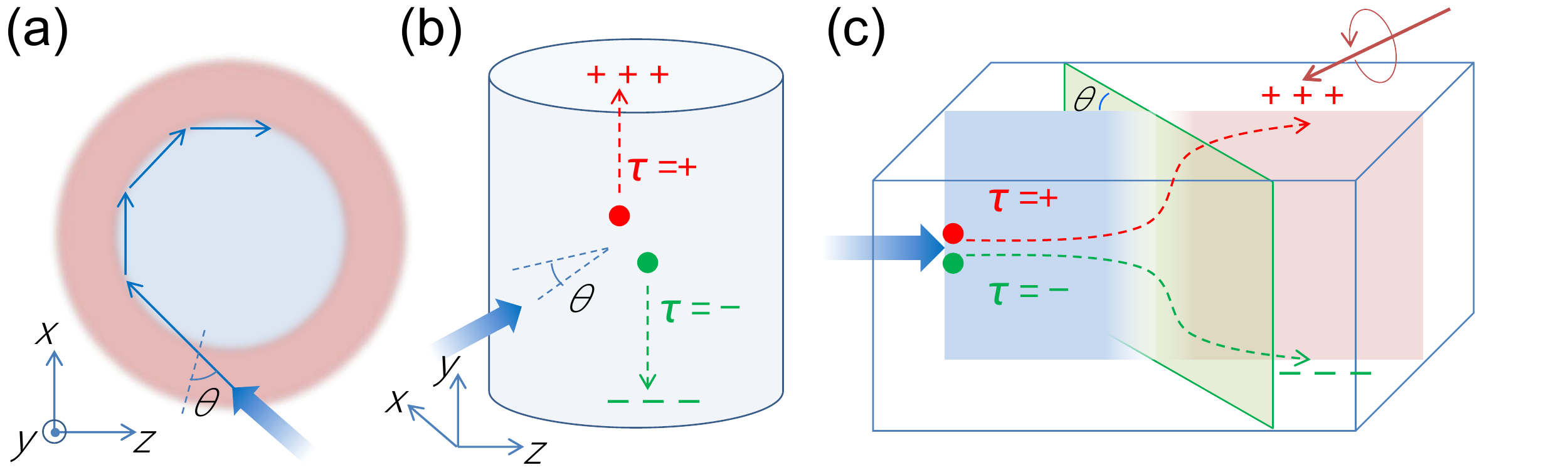}
\caption{(a) Top view of an electron undergoing multiple (total) reflections in a cylindric potential well of WSM.
(b) Side view of the enhanced CHE in (a).
(c) The chirality accumulation on the top and bottom surfaces in transmission through the green interface, which is detectable by the imbalanced absorbance of the left and right circularly polarized lights.}
\label{fig2}
\end{figure}

The CHE differs from the anomalous Hall effect (AHE) and the valley Hall effect (VHE). To view their distinctions, we now reproduce the Hall conductivities of WSM using the EOM.
An electric field produces a force $\dot{\bm k}=-e{\bm E}$, leading to an anomalous velocity $\dot{\bm r}=e{\bm E}\times{\bm{\mathit\Omega}}$.
From the current $I=-e\sum_{\bm k}\dot{\bm k}$, we derive that
\begin{eqnarray}\label{sigma-H}
\sigma_{ij}=\epsilon^{ij\ell}\frac{e^2}{4\pi\hbar}\int\frac{d k_\ell}{2\pi}\sgn(\tau v_\ell k_\ell)\;.
\end{eqnarray}
For the model~(\ref{Heff}) with two Weyl points at $\tau b\hat{z}$, $\sigma_{yz}$ and $\sigma_{zx}$ vanish
due to the cancelation of Hall contributions from the two chirality $\tau=\pm$, or from the opposite momenta $\pm k_\ell$.
Yet, $\sigma_{xy}$ survives because $\sgn(\tau v_z \tau b)=1$.
We can further obtain $\sigma_{xy}=be^2/\pi h$, assuming that the ``gaps'' $\tau v_z k_z$ are ``inverted'' at $0<\tau k_z<b$
and taking into account the lattice regularization of the linear bands. For the same reasons the VHE,
which requires an extra $\tau$-factor in the integrand of Eq.~(\ref{sigma-H}), simply vanishes.

Evidently, in the above case the AHE arises from $\mathit\Omega_z$
(i.e., $\mathit\Omega_z$ leading to $\sigma_{xy}=b e^2/\pi h$ and $\mathit\Omega_{x(y)}$ producing $\sigma_{yz(zx)}=0$),
whereas the CHE can arise from any component of $\bm{\mathit\Omega}$.
Our results highlight that $\bm{\mathit\Omega}$ is a two-form that has three components in the 3D case,
instead of one component like the 2D case.
In fact, the CHE has no counterpart in 2D, e.g., in graphene,
because the incident plane coincides with the atomic layer and a transverse shift normal to the layer is not well-defined.
But an AHE (VHE) is possible in graphene when an energy gap opens at its Weyl points with $\mathcal{T}$ ($\mathcal{P}$) symmetry breaking.

Experimentally, the adiabatical variation of Fermi energy and velocities may be controlled by local gates and strain effects, respectively.
In a cylindric potential well, the incident Weyl wave-packets undergo multiple reflections, leading to an enhanced CHE,
as shown in Fig.~\ref{fig2}(a) and Fig.~\ref{fig2}(b).
Because of the CHE in the multiple (total) reflections or in transmission,
the left- and right- handed Weyl fermions are separated toward opposite surfaces,
as sketched in Fig.~\ref{fig2}(c).
The surface chirality accumulation can be detected by the imbalanced absorbance of the left and right circularly polarized lights.
When $\mathcal{P}$ symmetry is further broken, the imbalance between the two chirality leads to a charge Hall effect with a voltage difference between the opposite surfaces.

Finally, our predictions can be generalized to a system with multiple pairs of Weyl points~\cite{Wan,Xu,Yang,ex2,ex3}.
Magnetic field effects can further be studied by including the Lorentz force $e{\bm B}\times\dot{\bm r}$
and the orbital magnetic moment $\bm m=e\bm{\mathit\Omega}\mathcal{E}$~\cite{XCN}, and second order corrections can also be incorporated into the EOM~\cite{Gao}.
Appealingly, the coupling $\bm E\cdot\bm B$ leads to the chiral anomaly~\cite{Son2},
whereas the coupling $-\bm m\cdot\bm B$ shifts the Weyl wave-packet energy depending on its chirality.
So far we have assumed that $\tau{\bm b}$, the Weyl point locations, is insensitive to the spatial variation of $V$ or $v$'s.
If $\tau{\bm b}$ does change notably, $\nabla\times{\bm b}({\bm r})$ would be a nontrivial axial gauge field,
and we would have to take into account the corresponding axial magnetic effect.

\begin{acknowledgements}
{\it Note added.}---After this work was finalized, a complementary and independent study~\cite{Xie} appeared,
with a similar CHE in reflection derived by a different approach.
\end{acknowledgements}

\bibliographystyle{apsrev4-1}

\end{document}